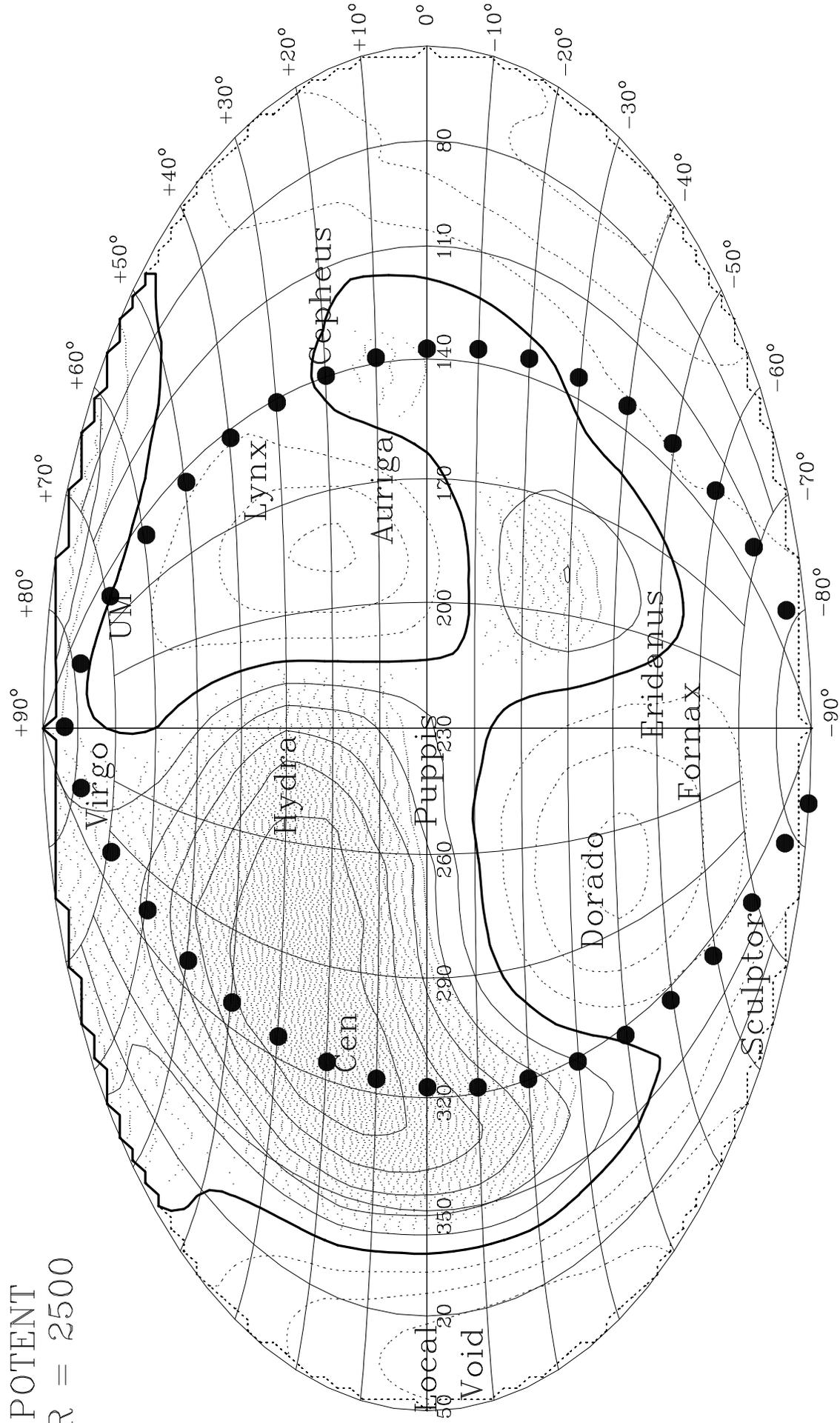

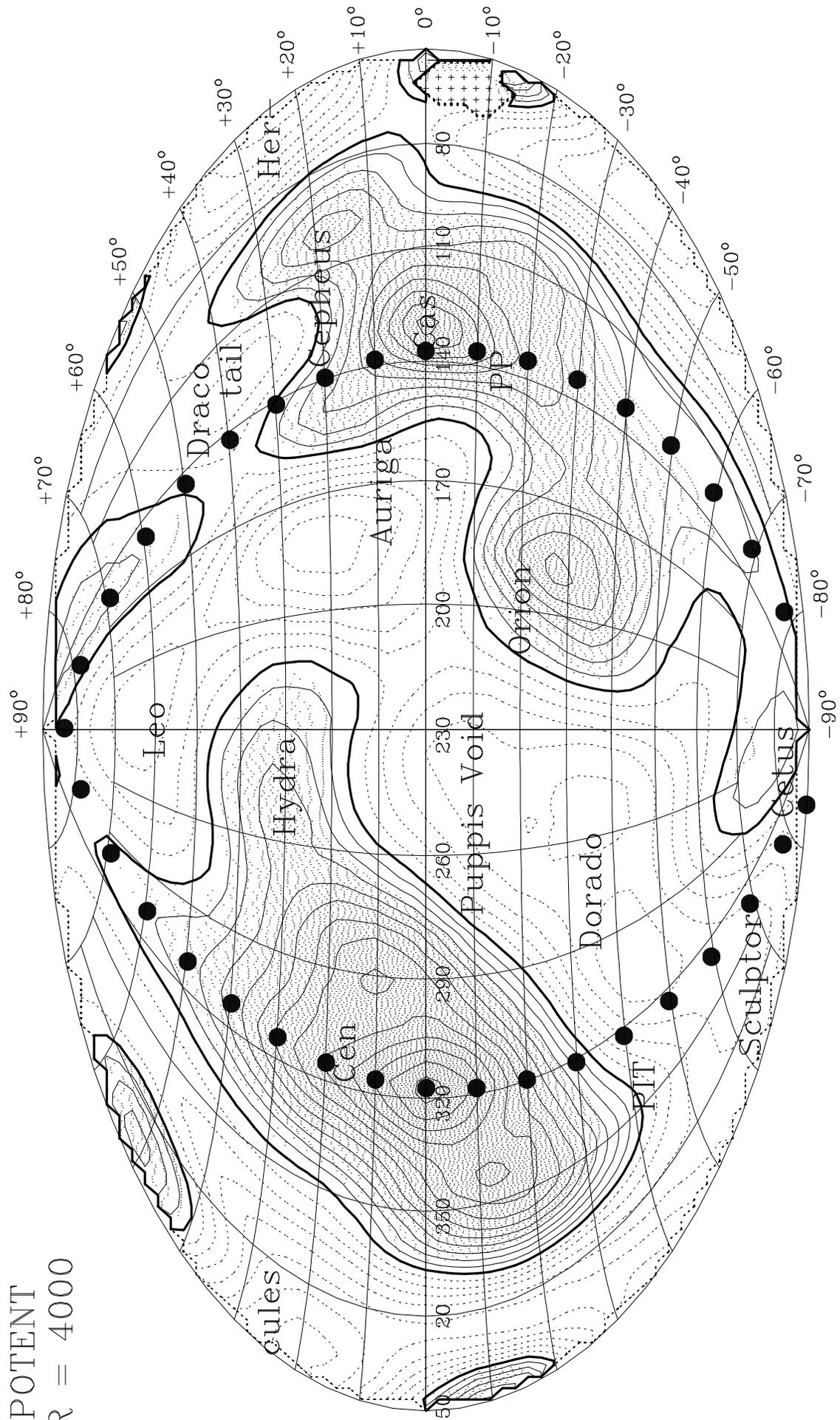

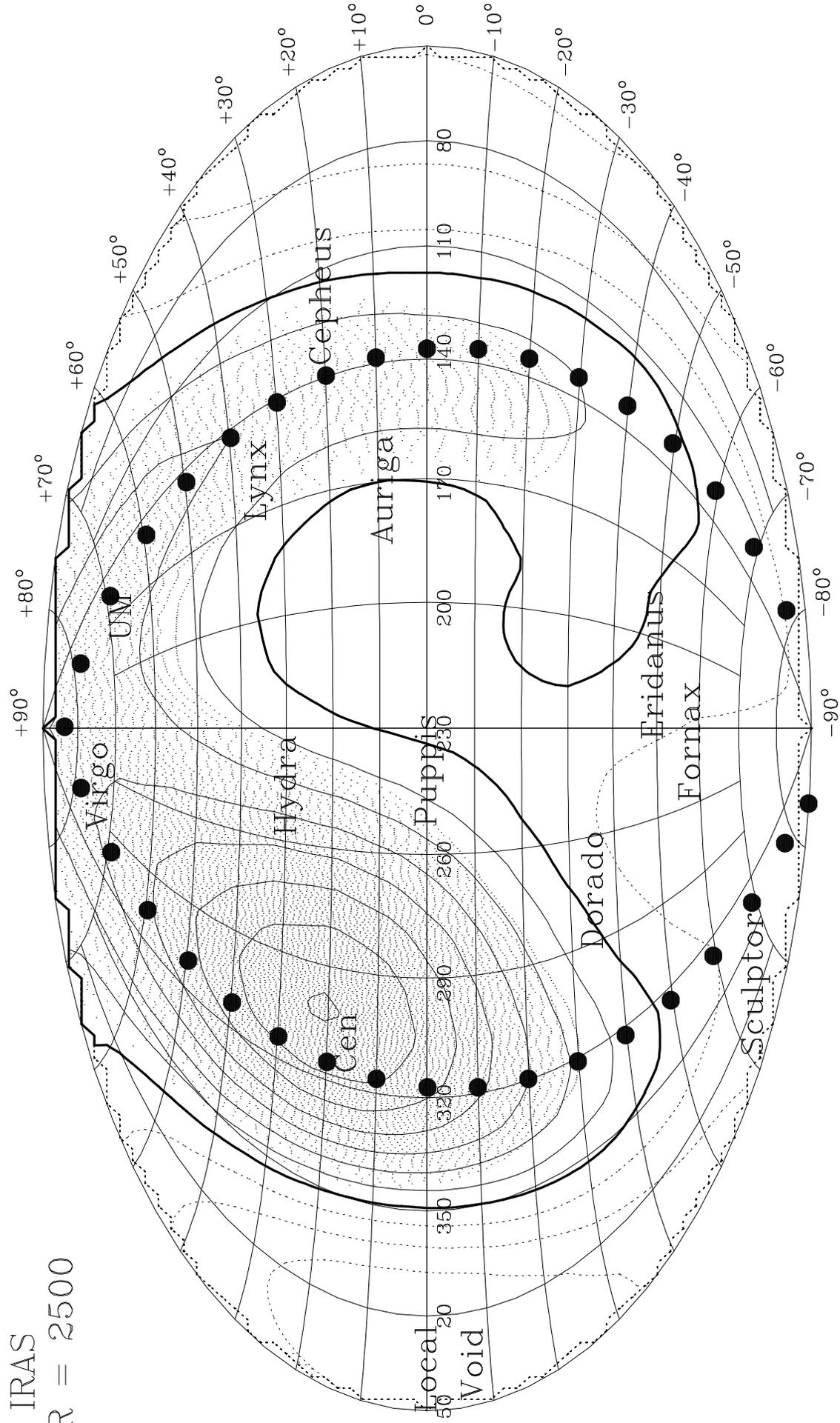

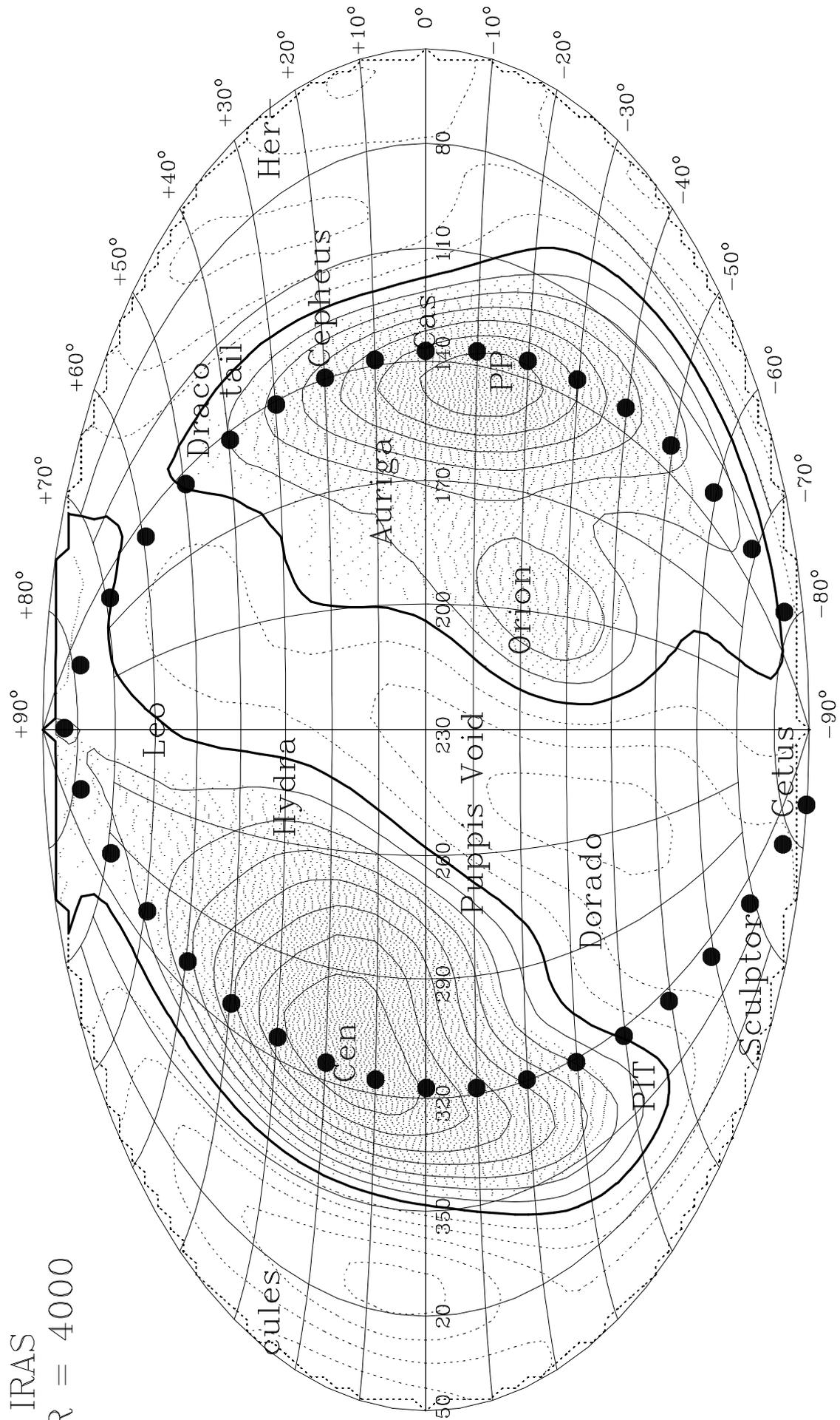

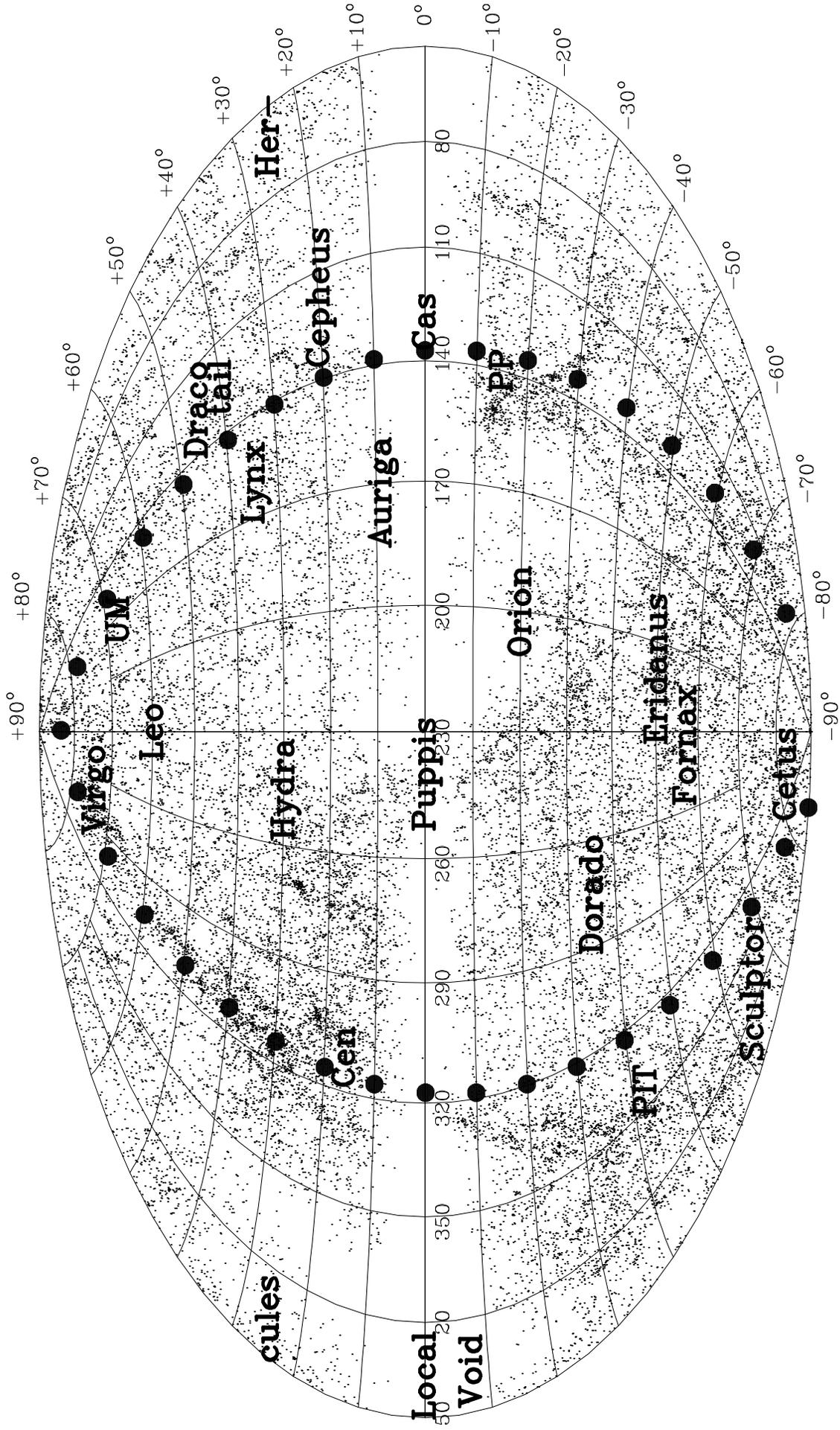

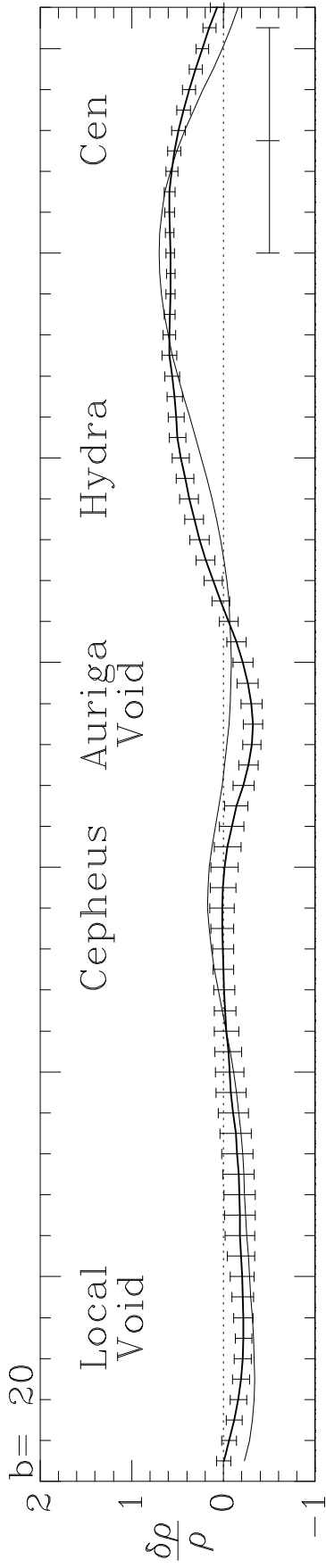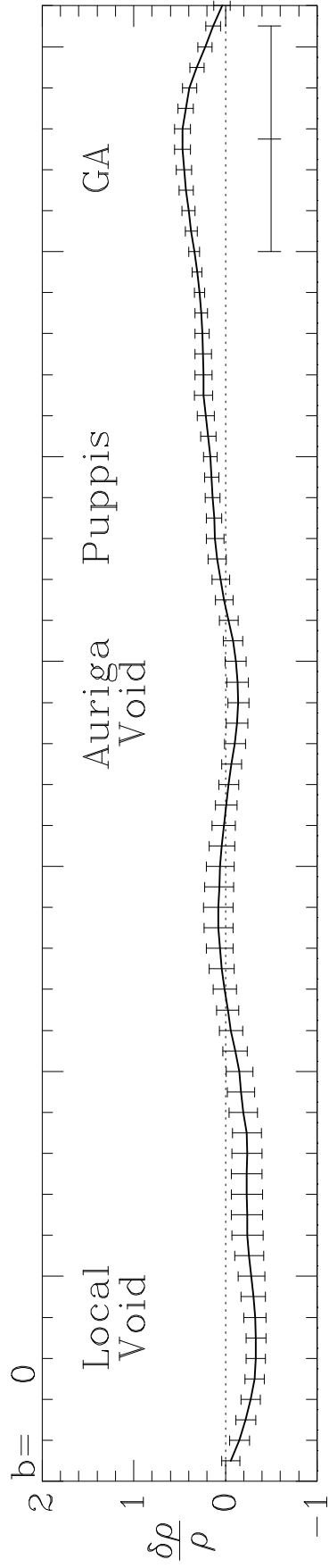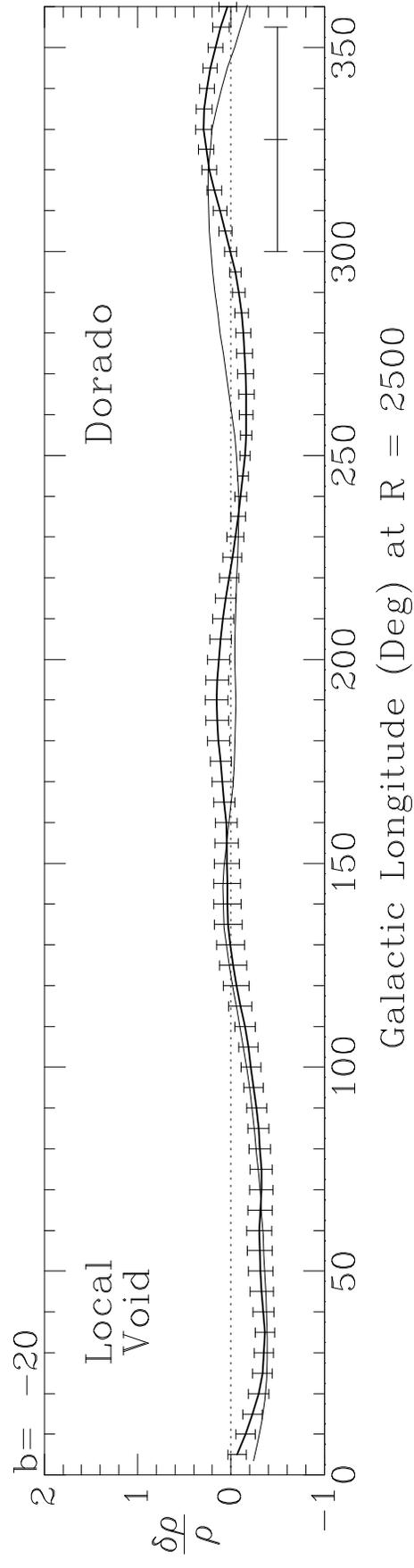

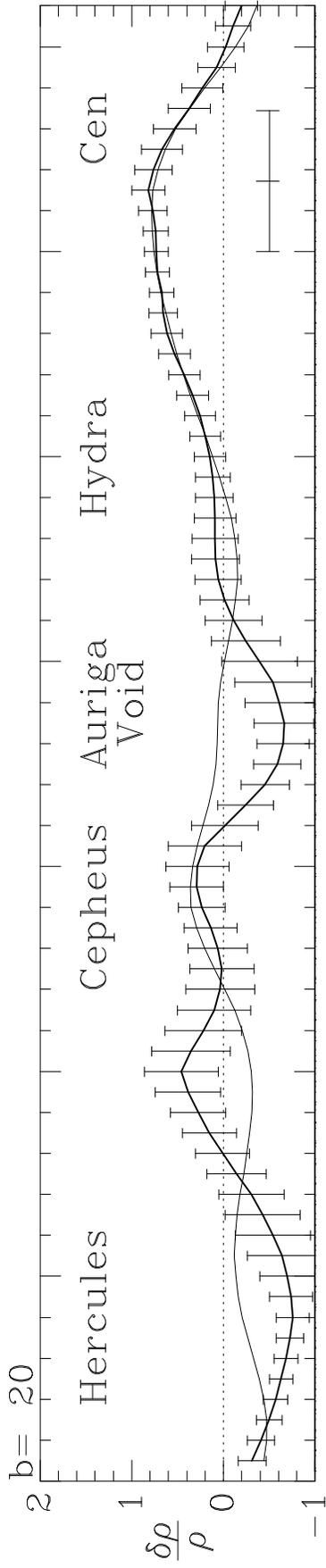
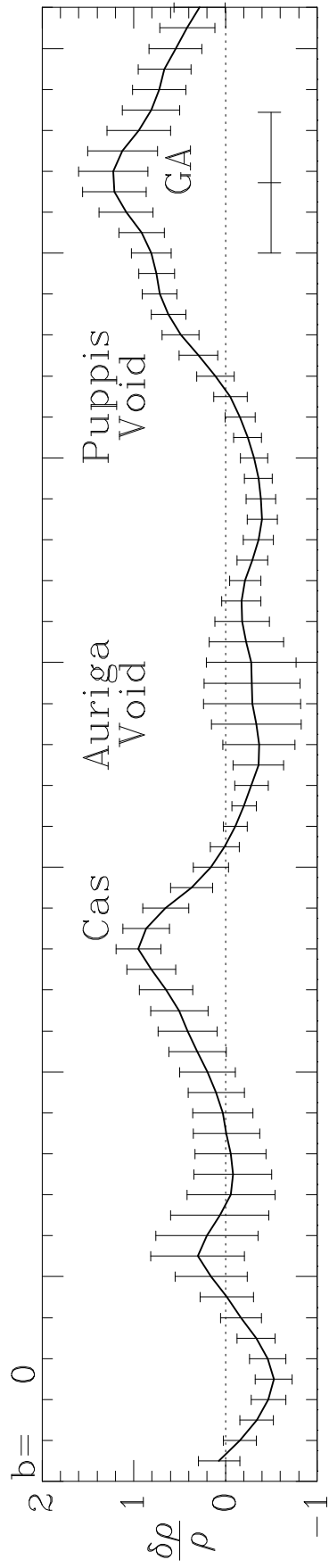
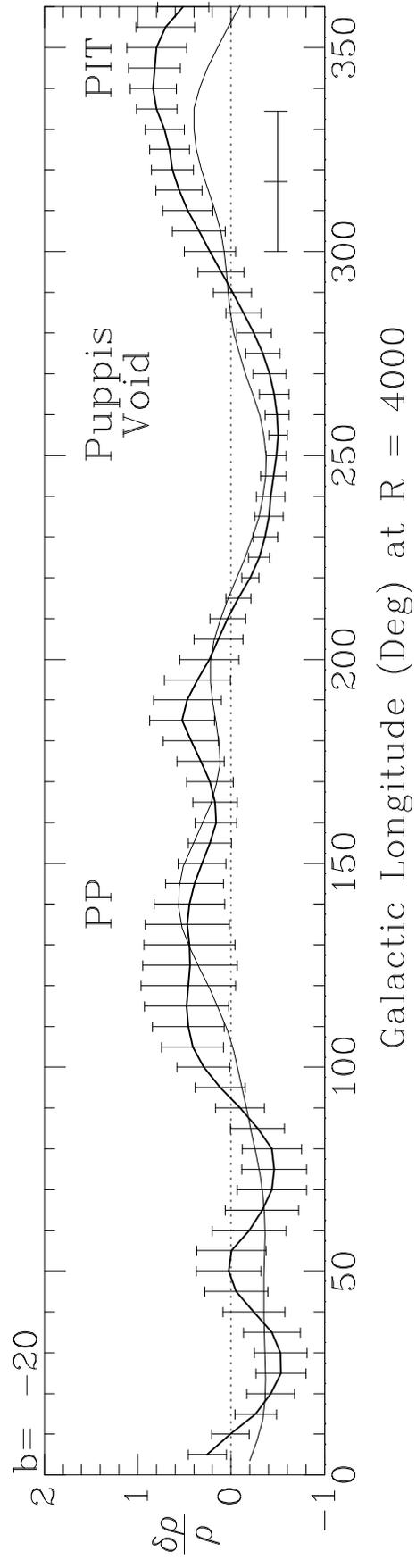

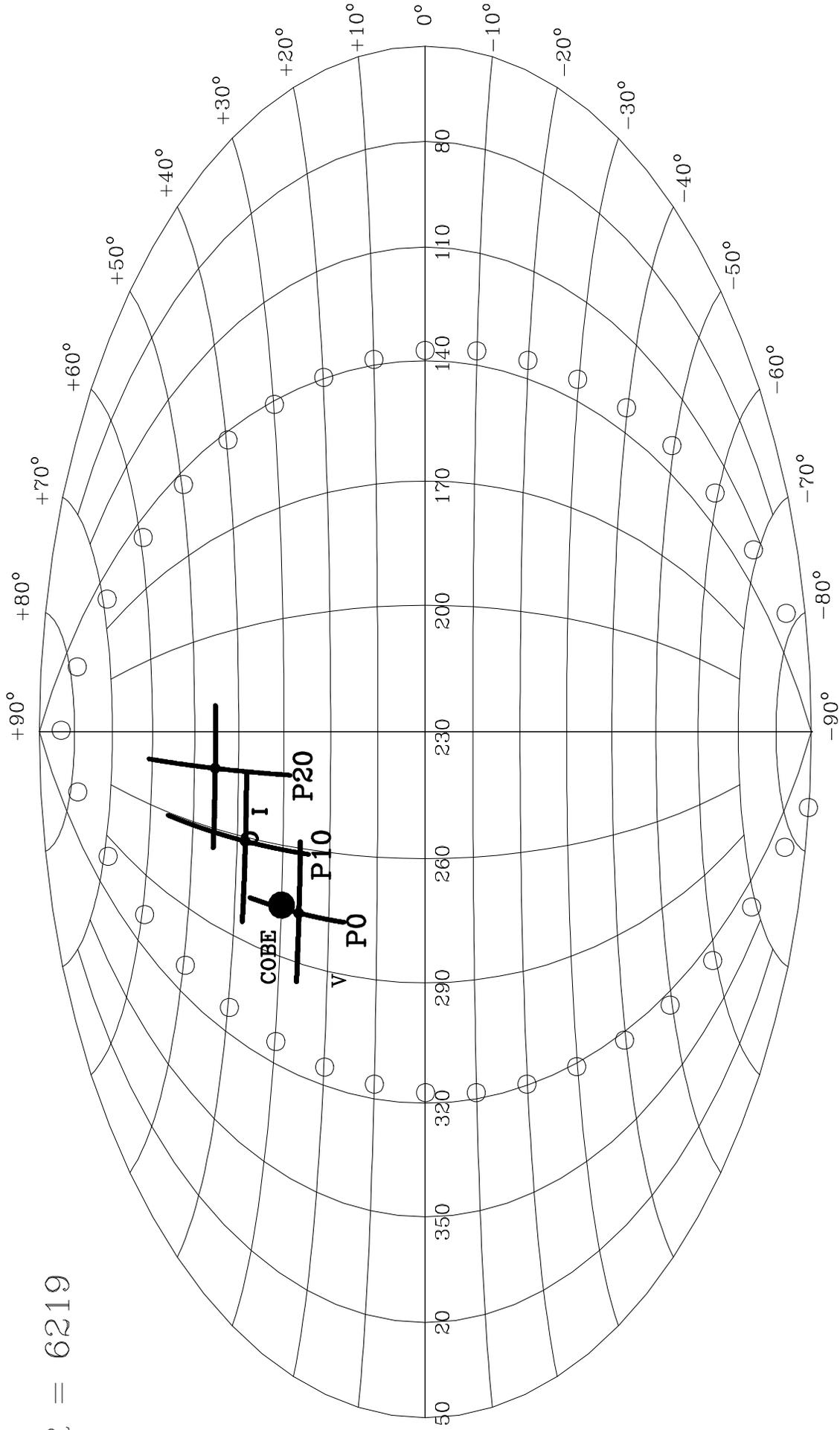

# LARGE-SCALE MASS DISTRIBUTION BEHIND THE GALACTIC PLANE

Tsafrir Kolatt[1], Avishai Dekel[1] and Ofer Lahav[2]

[1]Racah Institute of Physics, The Hebrew University, Jerusalem 91904, Israel
*tsafrir@astro.huji.ac.il   dekel@vms.huji.ac.il*
[2]Institute of Astronomy, Cambridge University, Madingley Road, Cambridge CB3 0HA, UK
*lahav@mail.ast.cam.ac.uk*

## ABSTRACT

We map the smoothed *mass*-density distribution in the Galactic zone of avoidance (ZOA), within 6000 km s$^{-1}$ of the Local Group, using POTENT reconstruction from peculiar velocities of galaxies. The interpolation into the ZOA is based on the assumed gravitational nature of the velocity field implying a potential flow. The main dynamical features at a distance $r \sim 4000$ km s$^{-1}$ are (a) the peak of the Great Attractor connecting Centaurus and Pavo at $l \simeq 330°$, (b) a moderate bridge connecting Perseus-Pisces and Cepheus at $l \simeq 140°$, and (c) an extension of a large void from the southern Galactic hemisphere into the ZOA near the direction of Puppis, $l \simeq 220° - 270°$. We find a strong correlation between the mass density and the IRAS and optical galaxy density at $b = \pm 20°$, which indicates that the main dynamical features in the ZOA should also be seen in galaxy surveys through the Galactic plane. The gravitational acceleration at the Local Group, based on the *mass* distribution out to $\sim 6000$ km s$^{-1}$, is strongly affected by the mass distribution in the ZOA: its direction changes by 31° when the $|b| < 20°$ ZOA is included, bringing it to within $4° \pm 19°$ of the CMB dipole.

*Key Words:* cosmology: observation — cosmology: theory — large-scale structure of the Universe — galaxies: clustering.

# 1 INTRODUCTION

The Galaxy obscures a strip of about one fifth of the optical extragalactic sky, and more than one tenth of the extragalactic sky observed by *IRAS*. The resulting incomplete sky coverage limits our understanding of several key issues of the large-scale structure. At the cosmographical level, it obscures the full structure in the Supergalactic Plane. At the dynamical level, it prevents us from fully exploring the origin of the motion of the Local Group (LG) with respect to the Cosmic Microwave Background (CMB). The sources of this motion are commonly estimated by summing up the contributions to the gravitational acceleration by *IRAS* galaxies (e.g. Rowan-Robinson *et al.* 1990, Strauss *et al.* 1992a) or by optical galaxies (e.g. Lynden-Bell, Lahav & Burstein 1989; Hudson 1993). The unknown mass distribution in the *ZOA* may therefore affect this estimate. For example, a newly discovered cluster in the *ZOA* towards Puppis ($l \approx 240°, |b| < 10°$, at a distance $R \approx 1500$ km/sec) may be responsible for a non-negligible fraction of the acceleration of the LG perpendicular to the Supergalactic plane (Lahav *et al.* 1993). The detection of new structures in the *ZOA* could also affect the peculiar velocities predicted from redshift surveys in regions away from the LG (e.g. Yahil *et al.* 1991). The comparison of velocities deduced from galaxy surveys with the measured CMB dipole or with observed peculiar velocities is used to estimate the ratio $\Omega^{0.6}/b$ of the density parameter $\Omega$ and the galaxy "biasing" parameter $b$ (see Dekel 1994 for a review). Hence, these estimates of important cosmological quantities are likely to be affected by the unknown contribution from the *ZOA*.

Recent studies have illustrated that our coverage of the *ZOA* can be much improved (see Kraan-Korteweg 1992 for a review). Currently, the exploration of the distribution of galaxies behind the *ZOA* is mainly done by searches for optical and *IRAS* galaxies in deep plates or catalogues, followed by redshift surveys, or by directed and "blind" searches in the 21 cm line of HI. One can also attempt a statistical approach to the problem, e.g. by expanding the observed incomplete sky in spherical harmonics and reconstructing the full sky harmonics using a Wiener filter which depends on a prior model and the noise level (Lahav *et al.* 1994).

A complementary approach presented here makes use of the peculiar velocities of galaxies, which are observed on the two sides of the *ZOA*, as probes of the mass distribution in the *ZOA* itself. The interpolation is based on the potential nature of a smoothed gravitational velocity field (Bertschinger & Dekel 1989), which allows a recovery of the transverse velocity components from the observed radial components, and then a reconstruction of the associated mass-density field. In other words, the peculiar velocity of a galaxy is due to the entire mass distribution, and hence it is related to the matter density field in neighboring regions which are not observed directly, such as the *ZOA*.

The mass-density field, smoothed on a large scale of 1200 km s$^{-1}$, is reconstructed by the *POTENT* method (§2). Whole sky density maps are shown (§3). They are compared with the *IRAS* galaxy density field in the two sides of the *ZOA*, to test for the correlation between the fields (§3). A similar correlation is expected to be valid in the *ZOA* as well. As a demonstration of the importance of the mass distribution in the *ZOA*, we compute the direction of the gravitational acceleration due to matter within 6000 km s$^{-1}$, with or without the *ZOA*(§4). These are compared to other relevant dipole directions. The results are discussed in §5.

## 2 POTENT RECONSTRUCTION

The POTENT method recovers the smoothed dynamical fluctuation fields of potential, velocity and mass density from observed radial peculiar velocities of galaxies, under quasilinear gravitational instability (Dekel *et al.* 1990; Bertschinger *et al.* 1990; Nusser *et al.* 1991; Dekel *et al.* 1994).

The raw data are distances, redshifts, and angular positions for a set of objects (galaxies or groups). The distances are estimated by the Tully-Fisher method for spirals (Aaronson *et al.* 1982) and the $D_n - \sigma$ relation for ellipticals and S0's (Lynden-Bell *et al.* 1988), with uncertainties of 15% and 21% respectively. The difference between the redshift and the estimated distance is the estimated radial peculiar velocity.

Given these sparsely-sampled radial velocities, POTENT first computes a *smoothed* radial-velocity field $u(\boldsymbol{x})$ in a spherical grid using a linear tensor window function which mimics a Gaussian of radius 1200 km s$^{-1}$ (Dekel *et al.* 1990; 1994). Weighting inversely by the local density near each object mimics equal-volume averaging which minimizes the bias due to sampling gradients. Weighting inversely by the distance variance of each object, $\sigma_i^2$, reduces the random effects of the distance errors.

The velocity field is recovered under the assumption of *potential* flow: $\boldsymbol{v}(\boldsymbol{x}) = -\boldsymbol{\nabla}\Phi(\boldsymbol{x})$. According to linear theory any vorticity mode decays in time as the universe expands, and based on Kelvin's circulation theorem the flow remains vorticity-free in the mildly-nonlinear regime as long as the flow is laminar. This has been shown to be a good approximation when collapsed regions are properly smoothed over (BD). The velocity potential can thus be calculated by integrating the radial velocity along radial rays,

$$\Phi(\boldsymbol{x}) = -\int_0^r u(r',\theta,\phi)dr' . \tag{1}$$

Differentiating $\Phi$ in the transverse directions recovers the two missing velocity components.

The underlying mass-density fluctuation field, $\delta(\boldsymbol{x})$, is then computed by

$$\delta(\boldsymbol{x}) = \left\| I - f(\Omega)^{-1}\frac{\partial \boldsymbol{v}}{\partial \boldsymbol{x}} \right\| - 1 , \tag{2}$$

where the bars denote the Jacobian determinant, $I$ is the unit matrix, and $f(\Omega) \equiv \dot{D}/D \simeq \Omega^{0.6}$ with $D(t)$ the linear growth factor (Peebles 1980). Eq. 2 is the NDBB solution to the continuity equation under the Zel'dovich assumption that particle displacements evolve in a universal rate (Zel'dovich 1970). This nonlinear approximation, which reduces to the familiar $\delta = -f(\Omega)^{-1}\boldsymbol{\nabla}\cdot\boldsymbol{v}$ in the linear regime, has been found to approximate the true density in N-body simulations with an *rms* error less than 0.1 over the range $-0.8 \leq \delta \leq 4.5$ (NDBB; Mancinelli *et al.* 1994).

The largest source of random uncertainty are the *distance errors*, $\sigma_i$, of the individual galaxies. The final errors are assessed by Monte-Carlo simulations, where the input distances are perturbed at random using a Gaussian of standard deviation $\sigma_i$, and each artificial sample is fed into POTENT. The standard deviation of the recovered $\delta$ at each grid point over the Monte-Carlo simulations, $\sigma_\delta$, serves as our estimate for the random distance error. In the well-sampled regions (out to 4000 km s$^{-1}$ in most directions and beyond 6000 km s$^{-1}$ in certain directions) the measurement errors are below 0.3, but they exceed unity in certain poorly sampled, noisy regions at large distances. To exclude noisy regions, we limit any quantitative analysis to points where $\sigma_\delta$ is smaller than a certain critical value. In this paper we limit the plots and the analysis to regions where $\sigma_\delta < 0.5$.

Our reconstructed fields are also subject to *biases*: the inhomogeneous Malmquist bias (IM) and a sampling gradient bias (SG). The IM bias is due to the coupling of distance errors and the clumpy distribution of galaxies from which the sample has been selected. Most of the IM has been removed from the raw data by heavy grouping in redshift space and by correcting the estimated distances using the *IRAS* density field (Fisher *et al.* 1992) as a tracer of the underlying galaxy density (Willick 1991; 1994; Dekel *et al.* 1994; Kolatt *et al.* 1994). The correction in $\delta$ is typically limited to $\sim 10 - 20\%$ reduction in density contrast. This error is small compared to the random errors.

The SG bias is the crucial issue concerning the interpolation into the *ZOA*. This bias arises from the inhomogeneous sampling of a spatially-varying velocity field within each smoothing window. We find, using Monte-Carlo simulations of $N$-body "data", that with our volume-weighting scheme this bias is reduced to be typically smaller than the scatter due to distance errors. But in empty regions which are large compared to the effective size of the smoothing window the SG bias might be severe enough to generate spurious flows. Such regions are excluded from any *quantitative* analysis by rejecting all grid points where the distance to the 4th nearest object, $R_4$, is greater than a certain critical value. With the Gaussian window of radius 1200 km s$^{-1}$ used here, we include only regions where $R_4 < 1500$ km s$^{-1}$, guaranteeing that there are at least a few objects within the effective volume of the window. This criterion excludes about one half of the *ZOA* at $r = 4000$ km s$^{-1}$, leaving us with only those regions where the sampling penetrates relatively deep into the *ZOA*.

The interpolation of *POTENT* across the *ZOA* is limited by the smoothing; The SG bias can be properly corrected as long as the half-width of the *ZOA* is comparable to the smoothing radius, or smaller. With 1200 km s$^{-1}$ Gaussian smoothing radius, corresponding to 17° at 4000 km s$^{-1}$, the interpolation into the Galactic plane must be severely biased at distances significantly larger than 4000 km s$^{-1}$.

The current sample of objects (galaxies or groups) with measured distances and redshifts has been carefully calibrated and put together from several data sets by Willick *et al.* (1994). The preliminary data used here is based on the earlier data compiled by Burstein (1990), and on recent spiral data by Mathewson *et al.* (1992), Willick (1991), Courteau (1992), Han and Mould (1990; 1992) and Mould *et al.* (1991). The sample used here has 2913 galaxies in 1264 objects.

# 3  MAPS OF MASS DENSITY

The reconstructed smoothed density fields are presented as whole-sky (Aitoff) area-preserving projections in Galactic coordinates at a given distance, with the Galactic plane along the equator and the south Supergalactic pole ($l \approx 230°$) at the center. Figure 1 shows the 1200 km s$^{-1}$-smoothed POTENT density field in shells at distances $R = 2500$ and 4000 km s$^{-1}$. The very small area where the POTENT reconstruction is not reliable ($\sigma_\delta > 0.55$) is marked out. Figure 2 shows in comparison the density field of IRAS galaxies brighter than 1.9 Jy (Strauss et al. 1992b), Gaussian smoothed at 1200 km s$^{-1}$ as in POTENT, in the same shells. The sky coverage of the IRAS sample is limited to 88% of the sky, with a ZOA at $|b| < 5°$. Figure 3 shows a projected optical map of UGC, ESO and MCG galaxies of apparent diameter larger than $\approx$1 arcmin (see e.g. Lahav 1987 and Lynden-Bell & Lahav 1988 for details). This optical map helps the orientation and the identification of named clusters.

Figure 4 focuses attention on the mass-density run along Galactic longitude at given Galactic latitudes: $b = 0°$ and $\pm 20°$. The IRAS density run is shown in comparison above and below the ZOA, at $b = \pm 20°$.

The most pronounced feature is the *Great Attractor* (GA). At $R = 4000$ km s$^{-1}$ it is a broad ramp of overdensity, covering the ZOA in the range $l = 270° - 360°$. It extends from the Hydra and Centaurus *high-density* regions in the north Galactic hemisphere to the Pavo-Indus-Telescopium (PIT) region in the south, and it actually peaks in the ZOA, with $\delta = 1.1 \pm 0.4$. The $R = 2500$ km s$^{-1}$ shell shows the near side of the GA with a similar extent and about one half the overdensity.

The region next to the GA, the range $l = 220° - 270°$ near Puppis, shows a *low-density* region in the 4000 km s$^{-1}$ shell, hidden behind an *overdensity* in the nearby shell. The *low density* region is an extension of the deep Sculptor void (Kauffman & Fairall 1991) between Sculptor and Dorado in the south Galactic hemisphere [The Sculptor void, by the way, has been used by Dekel & Rees (1994) for estimating $\Omega$]. The foreground mild overdensity extends all the way from Hydra to Eridanus, intersecting the ZOA at the vicinity of Puppis.

The range $l = 155° - 220°$ is recovered in the 4000 km s$^{-1}$ shell with relatively large uncertainty. However, it seems to be a low-density extension of the void near Auriga in the north Galactic hemisphere. The high-density near Orion in the south does not continue in the ZOA.

The interval $l = 120° - 155°$ near Cassiopeia (Cas) shows a *connection* between the Perseus-Pisces (PP) overdensity in the south Galactic hemisphere and the high-density region in Cepheus.

Finally, the interval $l = 0° - 60°$ is dominated by the *Local Void* (LV, see Tully & Fisher Atlas 1987). It extends across a large range of Galactic latitudes, with it's deepest point in the ZOA $\delta = -0.4 \pm 0.2$ in the nearby shell. This void extends deeper into the 4000 km s$^{-1}$ shell, though it is only poorly recovered in the ZOA itself. It's north galactic extension in the Hercules region is as deep as $\delta = -0.8 \pm 0.2$.

The general correlation between the POTENT mass density field and the galaxy density field of IRAS and of optical galaxies (Dekel et al. 1994; Hudson et al. 1994) indicates that the POTENT mass density can serve as a prediction for the galaxy distribution in the ZOA. The $b = \pm 20°$ panels of Fig. 4 demonstrate this correlation; the POTENT and IRAS curves agree within one standard deviation of the POTENT random error in most regions. In the 4000 km s$^{-1}$ shell the agreement is good in the GA, both in Centaurus and in most of PIT. It is also good in the Puppis void, and in PP and its northern extension near Cepheus. The densities agree everywhere in the 2500 km s$^{-1}$ shell.

The correlation between POTENT and IRAS is weak in two voids in the $b = 20°$ strip at 4000 km s$^{-1}$; the voids in Auriga and Hercules are significantly deeper in POTENT than in IRAS. These local failures of the general correlation warn us that the correlation is not perfect. Although the same main features show up in both maps, they do not always coincide in position, extent and amplitude.

The POTENT-IRAS correlation outside the ZOA indicates that the POTENT prediction in the ZOA should be a reasonable approximation for the galaxy distribution there.

## 4 THE GRAVITATIONAL ACCELERATION AT THE LOCAL GROUP

The unknown mass distribution hidden behind the ZOA could strongly affect the gravitational acceleration at the Local Group and the important conclusions implied by it. The POTENT density fluctuation field enables a computation of this dipole, which we do in detail and study the consequences of elsewhere (Kolatt & Dekel, in preparation). For the purpose of demonstrating the possible effect of the mass behind the ZOA, we bring here preliminary estimates of the direction of the gravitational dipole direction, with and without the ZOA, in comparison to other relevant dipole estimates.

In linear theory, the vectors of peculiar velocity and acceleration are aligned, such that the peculiar velocity at the origin is given by the integral over space (Peebles 1980):

$$V = \frac{H_0 \Omega^{0.6}}{4\pi} \int \frac{\hat{x}}{x^2} \delta(x) \, \mathrm{d}^3 x \; , \qquad (3)$$

where $\delta(x) \equiv (\rho(x) - \bar{\rho})/\bar{\rho}$ is the mass-density fluctuation field. The recovered POTENT mass density field allows one to compute this integral with a reasonable accuracy out to $\sim 6000$ km s$^{-1}$, free of assumptions about how galaxies trace mass. This is done using the POTENT values of $\delta$ on a grid, with too erroneous values being replaced by $\delta = 0$. Since there is no clear evidence that this integral converges within 6000 km s$^{-1}$ (e.g. Juszkiewicz, Vittorio & Wise 1990; Lahav, Kaiser & Hoffman 1990 and Strauss et al. 1992a), we are actually computing a limited approximation to the full acceleration. However, this is probably enough for demonstrating the role of the mass in the ZOA.

Figure 5 shows the direction of the gravitational acceleration, with $\delta$ set to zero in a ZOA defined alternatively by $b = \pm 0°, \pm 10°,$ and $\pm 20°$ (marked P0, P10, and P20 respectively). The vectors point at $(l, b) = (277° \pm 19°, +26° \pm 11°)$, $(261° \pm 21°, +38° \pm 16°)$, $(241° \pm 22°, +45° \pm 17°)$ respectively. The error bars are due to the random distance errors, as derived by the Monte-Carlo noise simulations The direction of the acceleration derived from the whole sky almost coincides (4°) with the direction of the LG velocity in the CMB frame [$(276°, 30°)$, Kogut et al. 1993, translated from the heliocentric CMB dipole measurements using the LG definition of Yahil et al. 1977]. The uncertainty due to the LG definition is of a few degrees (deVaculeurs et al. 1976). The directions deviates by 15° and 31° when a ZOA of ±10° and ±20° are cut out, respectively. The direction of P20 thus represents a 2-sigma deviation from the COBE dipole direction. The important role of the mass in the ZOA reflects the fact that big structures, such as the GA, peak near the Galactic plane.

It is of interest to compare our POTENT dipole direction to the dipoles derived from the galaxy distribution within 6000 km s$^{-1}$ in IRAS (e.g. Rowan-Robinson et al. 1990; Strauss et al. 1992a) and in optical catalogs (e.g. Lynden-Bell 1989; Hudson 1993). The IRAS direction shown in Figure 5 is (252°, 35°) (Strauss et al. 1992a, model b4), and the optical direction is (260°, 37°) (Hudson 1993, cloned mask model). The comparison is not straightforward because of the different selection effects, the different ways the ZOA was corrected for, and the possibility of non-trivial galaxy biasing which might affect the dipole direction. It is remarkable, however, that the galaxy dipoles are all pointing to within 25° of the CMB dipole, with the galaxy dipoles agreeing nicely with the P10 gravitational dipole from POTENT.

The POTENT velocity pointing at (290°, 17°) in Figure 5 is the 1200 km s$^{-1}$-smoothed velocity at the origin in the CMB frame, which is derived independently of the ≈ 6000 km s$^{-1}$ limit to the reliable recovery of POTENT density. It's deviation from P0 reflects (a) the local (anomaly) motion of the LG relative to the sphere of Gaussian radius 1200 km s$^{-1}$ about it, (b) possible nonlinear effects which violate the proportionality of acceleration and velocity, (c) the error made in evaluating P0 from the mass distribution within 6000 km s$^{-1}$, and (d) other errors in the POTENT analysis.

## 5 DISCUSSION

The potential nature of the large-scale velocity field, as predicted by the theory of gravitational instability, provides a method for interpolation into the ZOA from the observed radial peculiar velocities above and below it. We have used this method to produce maps of mass-density in the ZOA. The demonstrated similarity between the mass-density field and the IRAS galaxy-density field outside the ZOA argues that the mass-density distribution in the ZOA is an approximation to the galaxy distribution there. The maps presented here could therefore serve as a guide for galaxy searches in the ZOA.

The main features predicted by the dynamics in the ZOA are:

1. The Great Attractor, connecting Centaurus and PIT, is predicted to rise to its peak very close to the Galactic plane, in the range $l = 310° - 340°$, at a distance of ≈ 4000 km s$^{-1}$. This is a region of first-priority for galaxy search.

2. A moderate bridge is expected between PP in the south Galactic hemisphere and Cepheus in the north, near $l = 130° - 140°$, at a distance of 4000 km s$^{-1}$ and beyond, confirming the connectivity of the structure in the Supergalactic plane.

3. In the range $l = 220° - 270°$, near Puppis and Galactic-south of Hydra, the dynamics predicts a nearby high-density region (in the 2500 km s$^{-1}$ map) in front of a void (in the 4000 km s$^{-1}$ map).

4. Extended voids are predicted in the ranges $l = 0° - 60°$ and $l = 150° - 220°$, though with large uncertainty already at 4000 km s$^{-1}$.

The dynamical importance of the mass distribution in the ZOA is demonstrated by its relative contribution, within the inner 6000 km s$^{-1}$, to the direction of the gravitational acceleration at the Local Group. We find the mass distribution in the ZOA to be of crucial importance. The inclusion of the ±20° ZOA changes the direction of the acceleration vector by $\approx 35°$. This is mostly due to the high-contrast features in the ZOA: the GA and the nearby void in the opposite side of the sky. The fact that the acceleration determined from the whole sky converges to the direction of the CMB dipole indicates that the POTENT interpolation into the ZOA is reliable.

Thus, any determination of a dipole from a galaxy survey should be very cautious regarding how the ZOA is being dealt with. This may, in turn, strongly affect the estimates of the cosmological density parameter, $\Omega$. Reliable results based on the galaxy distribution require a deep and uniform galaxy search through the whole ZOA, at a wavelength where obscuration is negligible (e.g. at 2 $\mu m$) as close as possible to the Galactic plane. Unfortunately, this method, being based on galaxy count, would always suffer from variations in obscuration in the ZOA.

The *potential* interpolation demonstrated here is a more direct alternative, in the sense that it directly recovers the mass density, independently of the "biasing" relation between galaxies and mass, and in the sense that it uses a physical interpolation scheme. An improved recovery of the mass distribution in the ZOA requires more distance estimates at low Galactic latitudes, in particular at large distances. However, this method may suffer indirectly from the same variations in obscuration, via possible systematic dependencies of the distance indicators on obscuration, so caution is called for here as well.

## Acknowledgments

TK and AD thank their peculiar velocity and POTENT collaborators, E. Bertschinger, D. Burstein, S. Courteau, A. Dressler, S.M. Faber, J. Willick, and A. Yahil, for the preliminary POTENT results used in the current application. This research has been supported by BSF grants no. 89-00194 and 92-00355, and by a Basic Research grant of the Israeli Academy no. 462/92. OL thanks the Center for Microphysics and Cosmology for its hospitality at the Hebrew University where most of this work has been done.

# FIGURE CAPTIONS

**Figure 1. :** The mass-density fluctuation field by POTENT from peculiar velocity data in two shells: (a) at 2500 $\mathrm{km\,s^{-1}}$, and (b) at 4000 $\mathrm{km\,s^{-1}}$. The density is smoothed by a three-dimensional Gaussian of radius 1200 $\mathrm{km\,s^{-1}}$. Contour spacing is $\Delta\delta = 0.1$, with $\delta = 0$ as a heavy contour, $\delta > 0$ solid and $\delta < 0$ dashed. The whole sky are shown, Aitoff equal-area projected into Galactic coordinates, with the Galactic plane at the equator, and Galactic longitude shifted such that the south Supergalactic pole is at the center. The Supergalactic plane is marked by solid dots. Small regions where the POTENT recovery is unreliable ($\sigma_\delta > 0.55$) are marked out.

**Figure 2. :** The galaxy-density fluctuation field from the IRAS 1.9 Jy survey (by Yahil *et al.* 1991). Coordinates, smoothing and contours as in Fig. 1.

**Figure 3. :** The optical galaxies from the UGC, ESO and MCG catalogs projected on the sky in the same coordinate system as in Fig. 1.

**Figure 4. :** The smoothed-density run in Galactic longitude at fixed latitudes, $b = 0°, \pm 20°$, in two shells: (a) at 2500 $\mathrm{km\,s^{-1}}$, and (b) at 4000 $\mathrm{km\,s^{-1}}$. The POTENT mass-density is the heavy curve, with error bars due to distance errors determined by Monte-Carlo simulations. The IRAS galaxy density runs at $b = \pm 20°$ are shown as thin lines for comparison. The angle corresponding to smoothing scale of $\pm 1200$ $\mathrm{km\,s^{-1}}$ at the given distance are marked by horizontal bars.

**Figure 5. :** Gravitational acceleration dipole directions on the sky, in the galactic coordinates of Fig. 1. The measured velocity of the Local Group relative to the CMB frame (COBE, Kogut *et al.* 1993) is marked by a filled dot. The gravitational accelerations as computed from the mass distribution by POTENT are marked $P0, P10$, and $P20$, for an excluded ZOA of $0°, \pm 10°$, and $\pm 20°$ respectively. Shown for comparison are the directions of an IRAS dipole (I) (Strauss *et al.* 1992) and an optical dipole (O)(Hudson 1993), derived from galaxies within 6000 $\mathrm{km\,s^{-1}}$. Also shown is the POTENT smoothed velocity in the CMB frame at the LG (V). The Supergalactic plane is marked by open circles.